# VSX J074727.6+065050: a new WZ Sagittae star in Canis minor

Jeremy Shears, Steve Brady, Greg Bolt, Tut Campbell, Donald F. Collins, Lewis M. Cook, Timothy R. Crawford, Robert Koff, Tom Krajci, Jennie McCormick, Peter Nelson, Joseph Patterson, Pierre de Ponthière, Mike Potter, Robert Rea, George Roberts, Richard Sabo, Bart Staels and Tonny Vanmunster

**Abstract**

We present photometry of the first reported superoutburst of the dwarf nova VSX J074727.6+065050 during 2008 January and February. At its brightest the star reached magnitude 11.4 and this was followed by a slow decline at 0.09 mag/d for 19 days, corresponding to the plateau phase. There was then a rapid decline at 1.66 mag/d to a temporary minimum at magnitude 16.6 where it stayed for 2 to 3 days after which there were six remarkable echo outbursts before the star gradually faded back towards quiescence at ~magnitude 19.5. The overall outburst amplitude was at least 8 magnitudes and it lasted more than 80 days. During the plateau phase we observed common superhumps with $P_{sh}$ = 0.06070(6) d, but the period increased to $P_{sh}$ = 0.06151(5) d coinciding with the end of the plateau phase and the onset of the rapid decline. This corresponds to a continuous period change with $\dot{P}$ = +4.4(9) x $10^{-5}$. During the echo outbursts there was a superhump regime with $P_{sh}$ = 0.06088(49) d. Evidence is presented which is consistent with the star being a member of the WZ Sge family of dwarf novae.

**Introduction**

Dwarf novae are a type of cataclysmic variable star in which a cool main sequence secondary star loses mass to a white dwarf primary. Material from the secondary falls through the inner Lagrangian point and, because it carries substantial angular momentum, does not settle on the primary immediately but forms an accretion disc. From time-to-time, as material builds up in the disc, thermal instability drives the disc into a hotter, brighter state causing an outburst in which the star brightens by several magnitudes. Dwarf novae of the SU UMa family occasionally exhibit superoutbursts which last several times longer than normal outbursts and may be up to a magnitude brighter. During a superoutburst the light curve of a SU UMa star is characterized by superhumps. These are modulations which are a few percent longer than the orbital period [1]. They are thought to arise from the interaction of the secondary star orbit with a slowly precessing eccentric accretion disc. The eccentricity of the disc arises because a 3:1 resonance occurs between the secondary star orbit and the motion of matter in the outer part of the accretion disc.

WZ Sge stars are a sub-class of the SU UMa family which are more highly evolved systems. WZ Sge systems have very short orbital periods, long intervals between outbursts, typically decades, and exceptionally large outburst amplitudes, usually exceeding 6 magnitudes. For a full account of dwarf novae, and specifically SU UMa and WZ Sge stars, the reader is directed to reference 1.

**Detection of the outburst**

VSX J074727.6+065050 (hereafter "VSX 0747"; the VSX nomenclature comes from the Variable Star Index of the AAVSO) was discovered by K. Itagaki (Yamagata, Japan) as a new mag 12.3 object on 2008 Jan 19.546 at RA 07 h 47 min 27.64 s Dec +06 deg 50min 50.0 sec (J2000) [2]. Spectroscopic observations by H. Naito and S. Narusawa (Nishi-Harima Astronomical Observatory, Japan) on 2008 Jan 19.63 revealed it to be a dwarf nova in outburst with a brightness of U = 12.3 [3]. An image of VSX 0747 in outburst is shown in Figure 1. Chris Lloyd (Open University, UK) pointed out that the position is coincident with GSC 2.3 N88A015933 (V = 19.08 +/- 0.5) and USNO-B1.0 0968-0160072 (B = 19.60 and 19.44, R = 19.99) and suggested it might be a WZ Sge system because of the large outburst amplitude [4]. Based on these catalogued magnitudes we take the quiescent magnitude to be ~19.5.

**Outburst light curve**

The lower frame of Figure 2 shows the light curve of the outburst based on the authors' observations, data from the American Association of Variable Star Observers (AAVSO) International Database [5] and measurements from ASAS-3 (The All Sky Automated Survey) [6]. We shall frequently refer to dates in the truncated form JD = JD – 2454000.

The ASAS-3 observations show that VSX 0747 was first detected on JD 470 (2008 Jan 5) at V = 14.3, on the rise to outburst, some 14 days before Itagaki's detection. The maximum recorded brightness (V=11.4; ASAS-3) was on JD 473 (Jan 8) and the 19 days from that time until JD 492 corresponds to the *plateau phase* commonly seen in the early stages of a dwarf nova outburst. During the plateau phase there was a slow decline at a rate of 0.09 mag/d. This was followed by a *rapid decline* at 1.66 mag/d between JD 493 and 495. The star then remained at a *temporary minimum* of about 16.6 for two or three days following which there were six remarkable rebrightening events, or *echo outbursts*. Each echo outburst was separated by 3.5 to 5 days and the star reached mag 13.4 - 13.9 at each peak. The rate of decline from the echo outbursts was 1.3 mag/d; we could not determine the rate of rise due to lack of data during the rises. The sixth echo outburst concluded with a decline to magnitude 17 and over the next 20 days there was a further slow fade to magnitude 17.5. Even on JD 550, 80 days after the first ASAS-3 detection, the star had V = 18.5 (+/- 0.2), still well above the quiescence magnitude of ~19.5. Thus the outburst amplitude was at least 8 magnitudes.

**Time resolved photometry**

The authors conducted V-band and unfiltered (C, clear) time-series photometry of VSX 0747 using the instrumentation shown in Table 1 and according to the observation log in Table 2. There is an advantage to having observers, themselves members of the CBA (Center for Backyard Astrophysics), BAA and AAVSO, located at many longitudes in both the northern and southern hemispheres as it resulted in particularly good coverage of the outburst. In all cases raw images were flat-fielded and dark-subtracted, before being analysed using commercially available differential aperture photometry software against the comparison star sequence given in the AAVSO chart 1071xj [7]. Systematic differences were apparent between observers, which were at least in part due to the different CCD cameras and filters used. To overcome these differences the data were normalised by removing the mean and linear trends, as is common practise when the aim is to investigate periodic signals. Sections of the resulting de-trended light curves are shown in Figures 3 to 9, with each section covering 2 days of the outburst, which clearly show the presence of modulations.

**Detection of superhumps and measurement of superhump periods**

*Plateau and rapid decline (JD 486 to 496)*

By the time we started time-resolved photometry on JD 486, the outburst was already well advanced into the *plateau phase*. Photometry from this part of the plateau (JD 486 to 492) is presented in Figure 3, which shows the presence of superhumps confirming this to be a superoutburst. Initially the peak-to-peak superhump amplitude was 0.12 mag (JD 486), gradually increasing to 0.18 mag (JD 488) and subsequently reducing to 0.10 mag (JD 490 – 492). During the first part of the rapid decline (JD 493 to 494) the amplitude remained at 0.10 mag (Figure 4, top panel) and increased to about 0.15 mag later in the decline (Figure 4, bottom panel).

To study the superhump behaviour, we first extracted the times of each sufficiently well defined superhump maximum from the individual light curves according to the Kwee and van Woerden method [8] using the Peranso software [9]. Times of 64 superhump maxima were found and are shown in Table 3 along with the error estimates from the Kwee and van Woerden method. Following a preliminary assignment of superhump cycle numbers to these maxima, we obtained the following linear superhump maximum ephemeris for the interval JD 486 to 490 (covering the first 71 superhump cycles):

$$JD_{Max} = 2454486.62946(12) + 0.06070(6) \times E \qquad \textit{Equation 1}$$

The observed minus calculated (O–C) residuals for the whole of the outburst relative to this ephemeris are plotted in the top panel of Figure 2. This suggests that sometime after JD 492 the superhump period changed. By reference to the outburst lightcurve in the lower panel of Figure 2, the period change corresponded with the end of the plateau phase and the start of the rapid decline. Analysing the times of maximum from JD 493 and 494 (superhump cycle 115 to 122), at the beginning of the rapid decline, yielded a slightly different linear superhump maximum ephemeris:

$$JD_{Max} = 2454486.54103(18) + 0.06151(15) \times E \qquad \textit{Equation 2}$$

However, the baseline over which this analysis was conducted was necessarily rather short since, as noted above, the superhump maxima were only sufficiently well defined to enable measurements to be made during the initial part of the rapid decline.

Thus from the analysis of the superhumps during the early part of the outburst, we can see two slightly different superhump regimes. During section of the plateau phase that we observed, the superhump period, $P_{sh}$, was 0.06070(6) d and this subsequently increased to $P_{sh}$ = 0.06151(15) d during the rapid decline. However, the data are also consistent with a continuous change in period with $\dot{P} = +4.4(9) \times 10^{-5}$. Continuous period changes SU UMa systems are much more common than sudden or abrupt changes and thus we consider the continuous period changed to be more likely.

To confirm our measurements of $P_{sh}$, we carried out a period analysis of the data during the *plateau phase* (JD 486 to 492) using the Lomb-Scargle algorithm in Peranso. This gave the power spectrum in Figure 10 which has its highest peak at a period of 0.06079(21) d (other peaks are 1d aliases) and which we interpret as $P_{sh}$; this value is consistent with our earlier measurement from the times of superhump maxima. The superhump period error estimate is derived using the Schwarzenberg-Czerny method [10]. Several other statistical algorithms in Peranso gave the same value of $P_{sh}$. A phase diagram of the data from the plateau phase, folded on $P_{sh}$ = 0.06079 d is shown in Figure 11. This exhibits the typical profile of superhumps in which the rise to superhump maximum is faster that the decline.

In a similar manner we analysed the data from JD 493 to 496, corresponding to the full extent of the rapid decline. This yielded $P_{sh}$ = 0.06115(35) d (and its 1 day aliases; power spectrum not shown), which is again consistent with the result from analysing times of superhump maxima.

*Temporary minimum (JD 496 to 498)*

During the temporary minimum 0.15 mag superhumps were present, but again the maxima were only sufficiently well defined to allow one individual time of maximum to be measured, namely superhump cycle 186 (Figure 5). Instead we performed a Lomb-Scargle analysis on the data which resulted in a strong signal corresponding to $P_{sh}$ = 0.06103(27) d.

*Echo outbursts (JD 499 to 523)*

Light curves of the four best observed echo outbursts, two to five, are shown in Figures 6 to 9 respectively. Superhumps were apparent during the fade from the echo outbursts, although their amplitude during the fifth echo was very low indeed (~0.05 mag). We also note that the amplitude of the superhumps increased as the star faded from the third echo outburst (Figure 7); a similar effect was observed during the 2001 outburst of WZ Sge [11]. However analysis of superhump periods was a real challenge due to the underlying rapidly changing brightness. Nevertheless we were able to measure the times of a total of six superhump maxima: 2 during the third echo outburst (JD 508), 3 during the fourth echo outburst (JD 513) and one during the fifth echo outburst (JD516). The O-C residuals relative to the ephemeris in Equation 1 of each of these superhumps is approximately 0.5 cycles (Figure 2), or about half a superhump period different from the earlier superhumps. In some SU UMa systems phase-shifted superhumps are seen later in an outburst and are interpreted as *late* superhumps which differentiate them from the *common* superhumps regime operating earlier in the outburst [1]. However, given the lack of O-C data in the preceding period, we cannot rule out a gradual evolution of the superhump period rather than a phase change

Lomb-Scargle analysis of the combined data from the second to fifth echo outburst yielded a strong signal with $P_{sh}$ = 0.06088(49) d. This value is very similar to the common superhump period during the plateau phase.

**Searching for the orbital period**

Although we could find no evidence for an orbital hump superimposed on the underlying superhumps by visual inspection of the light curves, careful period analysis has in the past revealed subtle orbital signals in SU UMa systems during superoutburst. Hence we performed a Lomb-Scargle analysis on the complete data set and this revealed a strong signal at 0.06080(09) d which is consistent with the $P_{sh}$ measured during the plateau phase and during the echo outbursts. Removing $P_{sh}$ from the power spectrum (pre-whitening) left a small signal with P = 0.06061(11) d. Could this signal be the orbital period, $P_{orb}$? If it were, then the "period excess", $\varepsilon$, where $\varepsilon = (P_{sh} - P_{orb}) / P_{orb}$, would be 0.003. However, such value would be much smaller that that of the smallest known period excess WZ Sge system, EG Cnc, which has $\varepsilon$ = 0.0067 [12]. Whilst not impossible, it is unlikely, as it would imply a very low mass ratio, q, between the secondary and the white dwarf primary. Using Patterson's relationship $\varepsilon = 0.18q + 0.29q^2$ [12], a value of $\varepsilon$ = 0.003 implies q ~ 0.016, which is lower than any currently known dwarf nova. We therefore conclude that this signal is unlikely to be the orbital period and its origin remains a mystery.

**VSX 0747 as a WZ Sge system**

Based on the observational evidence obtained during the outburst of VSX 0747, we conclude that it is a WZ Sge star. The outburst amplitude of at least 8 magnitudes is higher than the maximum of about 6 magnitudes of most SU UMa dwarf novae. It is similar to several WZ Sge stars which typically have amplitudes in the range of ~ 6 to 8 magnitudes, including WZ Sge, HV Vir, EG Cnc and AL Com [11, 13, 14, 15]. The outburst light curve and superhump evolution is very similar to other WZ Sge systems including WZ Sge itself, AL Com and ASAS J002511+1217.2 [11, 15, 16]. However, it is the remarkable series of echo outbursts which were detected in VSX 0747 that are particularly suggestive of it being a WZ Sge system. Although echo outbursts are not completely diagnostic of WZ Sge status, since they have not been detected in all such stars and some SU UMa stars have also undergone rebrightening after a normal outburst, multiple echo outburst have been seen in several WZ Sge systems. For example, EG Cnc exhibited six echo outbursts during its 1996 outburst [14], SDSS J080434.20+510349.2 exhibited eleven [17] and WZ Sge itself exhibited twelve such echoes [11]. We note that the long duration of the outburst of VSX 0747 – it was still half a magnitude above quiescence 80 days after the first detection – is also consistent with a WZ Sge classification.

WZ Sge stars have very short orbital periods. Whilst our observations did not allow $P_{orb}$ to be measured, the value of $P_{sh}$ that we found places VSX 0747 only slightly above the period minimum (~78 min or 0.054 d [1]) in the distribution of orbital periods of SU UMa systems, where WZ Sge systems are common. Photometry at quiescence may reveal orbital modulations which would allow $P_{orb}$ to be determined, but this will be technically challenging considering the faintness of VSX 0747 and would require access to a large telescope. Although equally challenging, spectroscopy in quiescence would also aid in classification since WZ Sge systems are more highly evolved than the majority of SU UMa systems and their optical spectrum is dominated by a cool white dwarf, with no secondary star visible. As noted previously, spectroscopy was conducted by Naito and Narusawa at the beginning of the outburst [3], but the spectrum was dominated by the accretion disc which is the main light source during an outburst.

The superhump period of most SU UMa stars decreases during a superoutburst and is generally explained by the accretion disc shrinking. By contrast, several short period dwarf novae, including WZ Sge systems, show increasing superhump periods [13] in a similar manner to the increase that we have observed for VSX 0747. The value of $\dot{P}$ we obtained for VSX 0747 is consistent with the typical values for $5 \times 10^{-5}$ for similar period WZ Sge systems [18]. It has been proposed that the superhump period increases when the accretion disc expands beyond its 3:1 resonance radius. The reason why such an expansion occurs is not fully understood, although it appears to be associated with a low mass ratio and/or a low mass-transfer rate, as well as other factors which have yet to be determined [18].

The only significant feature which is often observed in WZ Sge stars, but that was not apparent from our observations of VSX 0747, is the presence of orbital humps: modulations in the light curve at the orbital period that sometimes occur before the appearance of superhumps. For example, WZ Sge exhibited orbital humps for the first 12 days of the 2001 outburst [11]. However, in the case of VSX 0747 our time-series unfortunately photometry missed the earlier stage of the outburst when orbital humps would have most likely made their appearance. Thus we must await a future outburst to find out whether such early orbital humps are present.

**Position of VSX 0747**

We measured the position of VSX 0747 on 8 images from Jan 24 at a time when it was near its brightest in the superhump cycle using the Astrometrica astrometry software [19] and the USNO-B1.0 catalogue. There was a variance of 0.2" in both RA and Dec in the 10 measurements. The mean position is:

07h 47m 27.73s +06d 50' 50.1"   (J2000)

Errors in both RA and Dec from Astrometrica/USNO-B1.0 are +/-0.3"

**Conclusion**

We present photometry of the first reported superoutburst of the dwarf nova VSX J074727.6+065050 during 2008 January and February. At its brightest the star reached magnitude 11.4 and this was followed by a slow decline at 0.09 mag/d for 19 days, corresponding to the plateau period. There was then a rapid decline at 1.66 mag/d to a temporary minimum at magnitude 16.6 where it stayed for 2 to 3 days after which there were six remarkable echo outbursts each separated by 3.5 to 5 days. The sixth echo outburst concluded with a rapid decline to magnitude 17 following which the star gradually faded back towards quiescence at magnitude ~19.5. The overall outburst amplitude was at least 8 magnitudes and it lasted more than 80 days.

Time resolved photometry revealed common superhumps during the plateau phase with a period of 0.06070(6) d, but this increased to 0.06151(5) d coinciding with the end of the plateau phase and the onset of the rapid decline. This corresponds to a continuous period change with $\dot{P} = +4.4(9) \times 10^{-5}$. During the echo outbursts the superhump period was 0.06088(49) d. We also noted that there was a 0.5 phase difference at this time compared with superhumps observed in the plateau phase. Whilst this could be indicative of late superhumps, given the lack of data in the preceding period we cannot rule out a simple period evolution.

Although we were not able to measure the orbital period, the very short superhump period is consistent with it being just above the period minimum in the orbital period distribution of dwarf novae. This observation, along with the large outburst amplitude, the longevity of the outburst and the echo outbursts, lead us to conclude that VSX 0747 is a member of the WZ Sge family of dwarf novae.

We urge observers, whether observing visually or with CCD cameras, to continue to monitor this fascinating star for future outbursts.

**Acknowledgements**

The authors gratefully acknowledge the use of observations from the AAVSO International Database contributed by observers worldwide. We thank Dr. Boris Gaensicke (University of Warwick, UK) for helpful discussions and the referees whose suggestions have improved the paper.

**Addresses:**
JS: "Pemberton", School Lane, Bunbury, Tarporley, Cheshire, CW6 9NR, UK [bunburyobservatory@hotmail.com]
SB: 5 Melba Drive, Hudson, NH 03051, USA [sbrady10@verizon.net]
GB: CBA Perth, 295 Camberwarra Drive, Craigie, Western Australia 6025, Australia [gbolt@iinet.net.au]
TC: 7021 Whispering Pine, Harrison, AR 72601, USA [jmontecamp@yahoo.com]
DFC: Warren Wilson College, Asheville, NC 28815, USA [dcollins@warren-wilson.edu]
LMC: CBA Pahala, Hawaii, USA [lcoo@yahoo.com]
TRC: Arch Cape Observatory, Oregon, USA [tcarchcape@yahoo.com]
RK: 980 Antelope Drive West, Bennett, CO 80102, USA [bob@AntelopeHillsObservatory.org]
TK: CBA New Mexico, PO Box 1351 Cloudcroft, New Mexico 88317, USA [tom_krajci@tularosa.net]
JM: CBA Pakuranga, Farm Cove Observatory, 2/24 Rapallo Place, Farm Cove, Pakuranga, Auckland, New Zealand [farmcoveobsextra.co.nz]
PN: Ellinbank Observatory, Victoria, Australia [pnelson@dcsi.net.au]
JP : Department of Astronomy, Columbia University, 550 West 20$^{th}$ Street, New York, NY 10027, USA [jop@astro.Columbia.edu]
PdP: CBA Lesve, Lesve-Profondeville, Belgium [pierredeponthiere@gmail.com]
MP : 3206 Overland Ave, Baltimore, MD 21214 USA [mike@orionsound.com]
RR : CBA Nelson, 8 Regent Lane, Richmond, Nelson, New Zealand [reamarsh@ihug.co.nz]
GR : 2007 Cedarmont Dr., Franklin, TN 37067, USA, [georgeroberts@comcast.net]
RS : 2336 Trailcrest Dr., Bozeman, MT 59718, USA [richard@theglobal.net]
BS: Alan Guth Observatory, Koningshofbaan 51, Hofstade, Aalst, Belgium [staels.bart.bvba@pandora.be]
TV: Center for Backyard Astrophysics Belgium, Walhostraat 1A, B-3401 Landen, Belgium [tonny.vanmunster@cbabelgium.com]

| Observer | Telescope | CCD | Filter |
|---|---|---|---|
| BS | 0.28 m SCT | Starlight Xpress MX716 | C |
| DFC | 0.2 m SCT | SBIG ST-7X | C |
| GB | 0.25 m SCT | SBIG ST-7 | C |
| JM | 0.35 m SCT | SBIG ST-8XME | C |
| JS | 0.1 m fluorite refractor | Starlight Xpress SXV-M7 | C |
| LC | 0.44 m reflector | Starlight Xpress SXV-H9 | C |
| MP | 0.35 m SCT | SBIG ST-8XME | C |
| PdP | 0.2 m SCT | SBIG ST-7XEI | C |
| PN | 0.32 m reflector | SBIG ST-8XE | C |
| RK | 0.25 m SCT | Apogee AP-47 | C |
| RR | 0.35 m SCT | SBIG ST-9 | C |
| RS | 0.25 m SCT | SBIG ST-10XME | V |
| SB | 0.4 m reflector | SBIG ST-8XME | C |
| TC* | 0.2 m SCT | SBIG ST-6b | C |
| TRC | 0.32 m SCT | SBIG ST-9XE | V |
| TK | 0.28 m SCT | SBIG ST-7E | C |
| TV | 0.35 m SCT | SBIG ST-7MXE | C |

**Table 1: Instrumentation used**
* TC's raw photometry data were reduced by GR

| Date in 2008 (UT) | Start time (JD-2454000) | Duration (h) | Observer |
|---|---|---|---|
| Jan 21 | 486.612 | 5.5 | MP |
| Jan 21 | 486.959 | 3.0 | PN |
| Jan 22 | 487.572 | 6.2 | TK |
| Jan 22 | 487.916 | 2.3 | RR |
| Jan 22 | 488.047 | 3.2 | GB |
| Jan 22 | 488.401 | 3.4 | PdP |
| Jan 23 | 488.572 | 2.5 | TK |
| Jan 23 | 488.768 | 4.4 | LMC |
| Jan 23 | 488.907 | 4.5 | RR |
| Jan 23 | 489.036 | 7.0 | GB |
| Jan 24 | 489.622 | 7.1 | RK |
| Jan 24 | 489.661 | 6.3 | RS |
| Jan 24 | 489.719 | 3.5 | TRC |
| Jan 24 | 489.904 | 3.6 | RR |
| Jan 24 | 490.312 | 3.9 | JS |
| Jan 25 | 490.612 | 7.3 | RK |
| Jan 25 | 490.677 | 3.8 | SB |
| Jan 25 | 490.756 | 5.6 | LMC |
| Jan 26 | 491.557 | 2.7 | MP |
| Jan 26 | 491.581 | 6.0 | SB |
| Jan 26 | 491.609 | 7.4 | RK |
| Jan 26 | 491.762 | 3.8 | LMC |
| Jan 26 | 491.806 | 3.4 | RS |
| Jan 26 | 491.879 | 5.2 | JM |
| Jan 26 | 492.355 | 1.8 | JS |
| Jan 27 | 492.578 | 4.7 | DFC |
| Jan 27 | 492.607 | 7.4 | RK |
| Jan 27 | 492.674 | 6.3 | TC |
| Jan 27 | 492.797 | 4.0 | LMC |
| Jan 28 | 493.559 | 5.6 | MP |
| Jan 28 | 493.911 | 1.3 | RR |
| Jan 28 | 493.963 | 2.9 | PN |
| Jan 29 | 494.903 | 3.5 | RR |
| Jan 30 | 495.901 | 2.6 | RR |
| Jan 31 | 496.905 | 2.7 | RR |
| Feb 1 | 497.911 | 2.8 | RR |
| Feb 3 | 499.883 | 4.0 | RR |
| Feb 7 | 504.393 | 2.9 | PdP |
| Feb 8 | 504.714 | 5.3 | TK |
| Feb 9 | 506.357 | 3.1 | JS |
| Feb 10 | 507.291 | 3.9 | JS |
| Feb 10 | 507.352 | 4.3 | BS |
| Feb 10 | 507.387 | 4.6 | TV |
| Feb 11 | 507.559 | 8.8 | TK |
| Feb 11 | 508.313 | 3.9 | BS |
| Feb 11 | 508.365 | 5.6 | TV |
| Feb 12 | 508.562 | 8.6 | TK |
| Feb 12 | 509.317 | 4.9 | TV |
| Feb 15 | 511.511 | 6.2 | MP |
| Feb 16 | 513.248 | 6.8 | TV |
| Feb 20 | 516.516 | 5.0 | MP |
| Feb 21 | 517.595 | 3.3 | MP |

**Table 2: Log of time-series observations**

| Superhump cycle number | Time of maximum (JD-2454000) | Error (d) | O-C [a] (cycles) |
|---|---|---|---|
| 0 | 486.63083 | 0.00141 | 0.02252 |
| 1 | 486.68979 | 0.00118 | -0.00606 |
| 2 | 486.75169 | 0.00093 | 0.01380 |
| 3 | 486.81076 | 0.00157 | -0.01297 |
| 7 | 487.05513 | 0.00071 | 0.01325 |
| 16 | 487.60046 | 0.00115 | -0.00195 |
| 17 | 487.65973 | 0.00055 | -0.02542 |
| 22 | 487.96194 | 0.00154 | -0.04624 |
| 24 | 488.08711 | 0.00176 | 0.01605 |
| 25 | 488.14731 | 0.00119 | 0.00790 |
| 36 | 488.81619 | 0.00192 | 0.02830 |
| 37 | 488.87879 | 0.00143 | 0.05969 |
| 38 | 488.93422 | 0.00095 | -0.02705 |
| 38 | 488.93677 | 0.00117 | 0.01497 |
| 39 | 489.00022 | 0.00161 | 0.06036 |
| 39 | 488.99625 | 0.00086 | -0.00505 |
| 40 | 489.06240 | 0.00134 | 0.08488 |
| 41 | 489.11583 | 0.00190 | -0.03486 |
| 42 | 489.17539 | 0.00096 | -0.05355 |
| 43 | 489.23667 | 0.00150 | -0.04391 |
| 44 | 489.29965 | 0.00170 | -0.00626 |
| 50 | 489.66458 | 0.00190 | 0.00629 |
| 51 | 489.72277 | 0.00111 | -0.03498 |
| 52 | 489.78454 | 0.00162 | -0.01726 |
| 53 | 489.84672 | 0.00157 | 0.00721 |
| 55 | 489.96602 | 0.00094 | -0.02721 |
| 56 | 490.02333 | 0.00129 | -0.08298 |
| 66 | 490.63372 | 0.00142 | -0.02625 |
| 67 | 490.69963 | 0.00161 | 0.05968 |
| 67 | 490.69683 | 0.00130 | 0.01354 |
| 68 | 490.76030 | 0.00137 | 0.05927 |
| 68 | 490.75604 | 0.00124 | -0.01092 |
| 69 | 490.81634 | 0.00167 | -0.01742 |
| 69 | 490.81467 | 0.00173 | -0.04494 |
| 70 | 490.88142 | 0.00163 | 0.05483 |
| 70 | 490.87361 | 0.00096 | -0.07385 |
| 71 | 490.93704 | 0.00120 | -0.02878 |
| 82 | 491.60500 | 0.00089 | -0.02354 |
| 83 | 491.66674 | 0.00151 | -0.00631 |
| 84 | 491.72740 | 0.00115 | -0.00689 |
| 86 | 491.85380 | 0.00142 | 0.07567 |
| 86 | 491.84739 | 0.00087 | -0.02994 |
| 87 | 491.90708 | 0.00130 | -0.04649 |
| 89 | 492.03090 | 0.00127 | -0.00645 |
| 100 | 492.69840 | 0.00176 | -0.00878 |
| 100 | 492.69627 | 0.00153 | -0.04388 |
| 101 | 492.76453 | 0.00153 | 0.08077 |
| 102 | 492.82345 | 0.00136 | 0.05153 |
| 102 | 492.82286 | 0.00115 | 0.04181 |
| 103 | 492.87914 | 0.00149 | -0.03093 |
| 103 | 492.88364 | 0.00090 | 0.04321 |
| 104 | 492.94403 | 0.00130 | 0.03819 |
| 115 | 493.61511 | 0.00156 | 0.09484 |
| 116 | 493.67791 | 0.00165 | 0.12953 |
| 117 | 493.73497 | 0.00118 | 0.06964 |
| 121 | 493.98477 | 0.00122 | 0.18532 |
| 122 | 494.04499 | 0.00131 | 0.17750 |
| 186 | 497.93572 | 0.00463 | 0.28079 |
| 362 | 508.63333 | 0.00270 | 0.53358 |
| 363 | 508.69344 | 0.00199 | 0.52395 |
| 439 | 513.30529 | 0.00223 | 0.50834 |
| 440 | 513.36556 | 0.00251 | 0.50134 |
| 441 | 513.42687 | 0.00280 | 0.51148 |
| 493 | 516.58669 | 0.00215 | 0.57237 |

**Table 3: Times of superhump maximum**
[a] O-C relative to the ephemeris in Equation 1

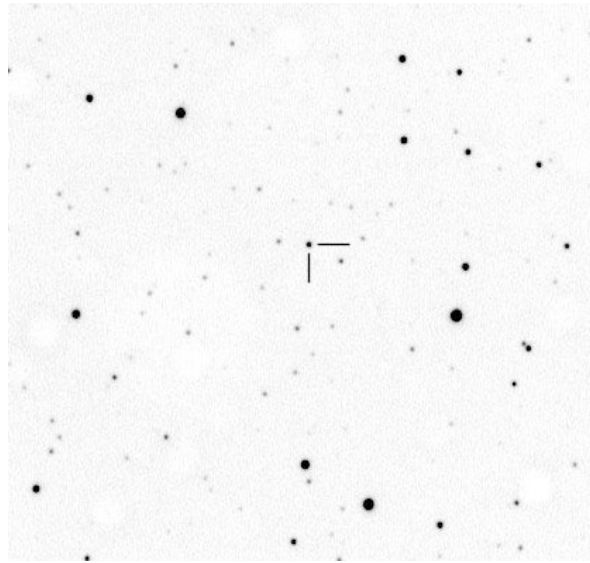

**Figure 1: VSX 0747 on 2008 January 24, 19.05 UT**
Field width 15 arcmin, N at top, W to right (Jeremy Shears)

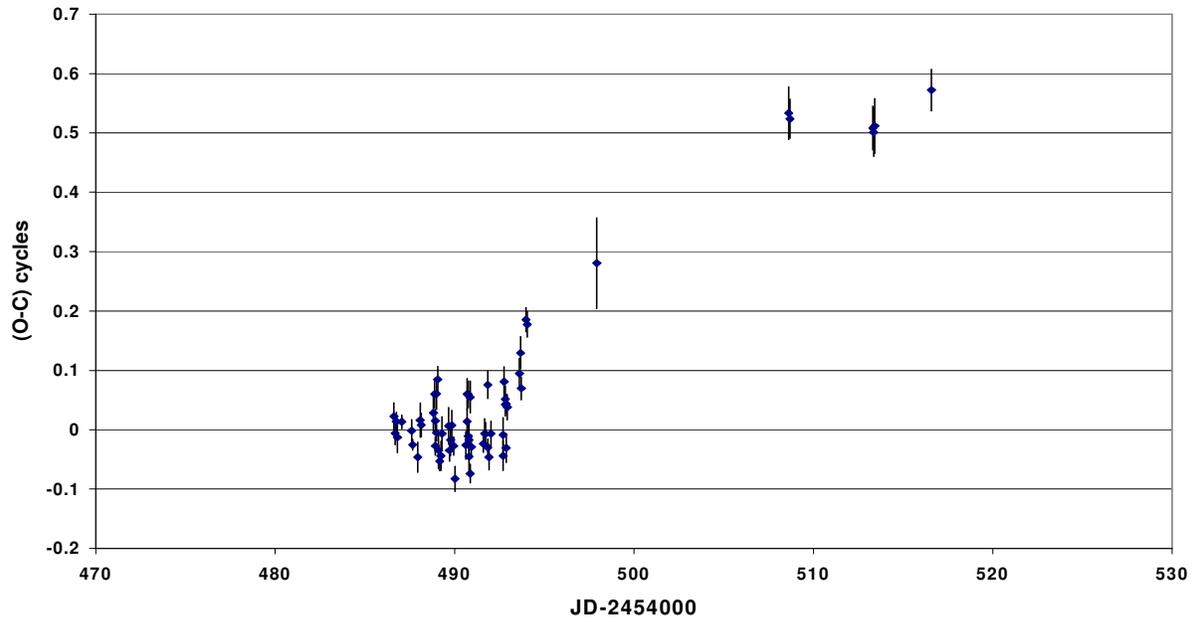

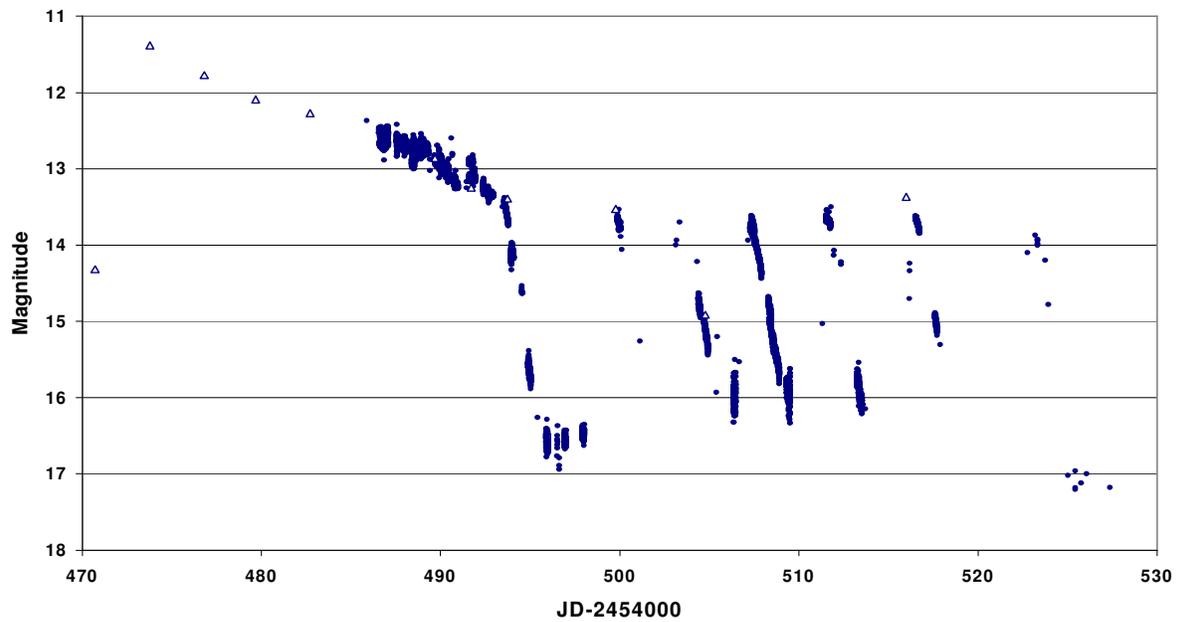

**Figure 2: Light curve of the outburst (bottom) and O-C diagram of superhump maxima (top)**

Light curve data are from the authors and the AAVSO International Database except for the triangles which are from ASAS-3. Error bars in the O-C diagram take into account the uncertainties in determining the times of superhump maximum show in Table 3.

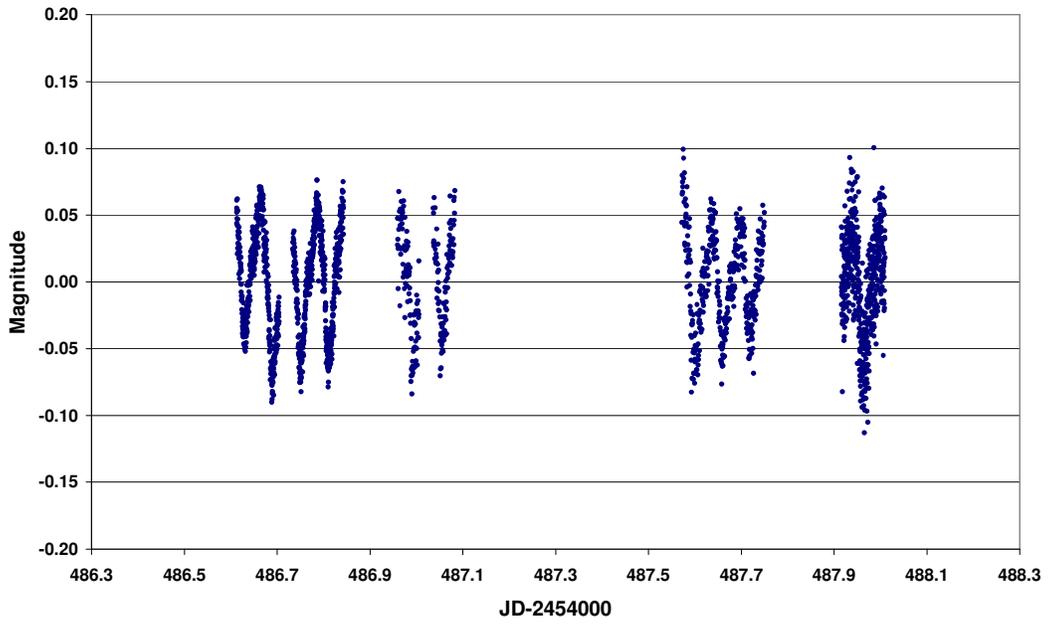
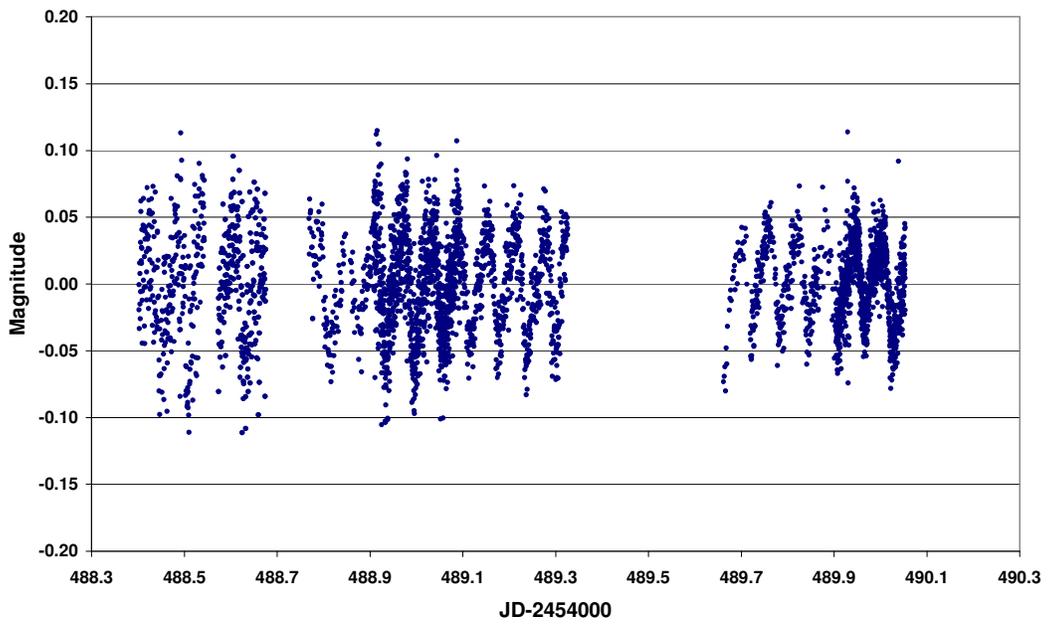

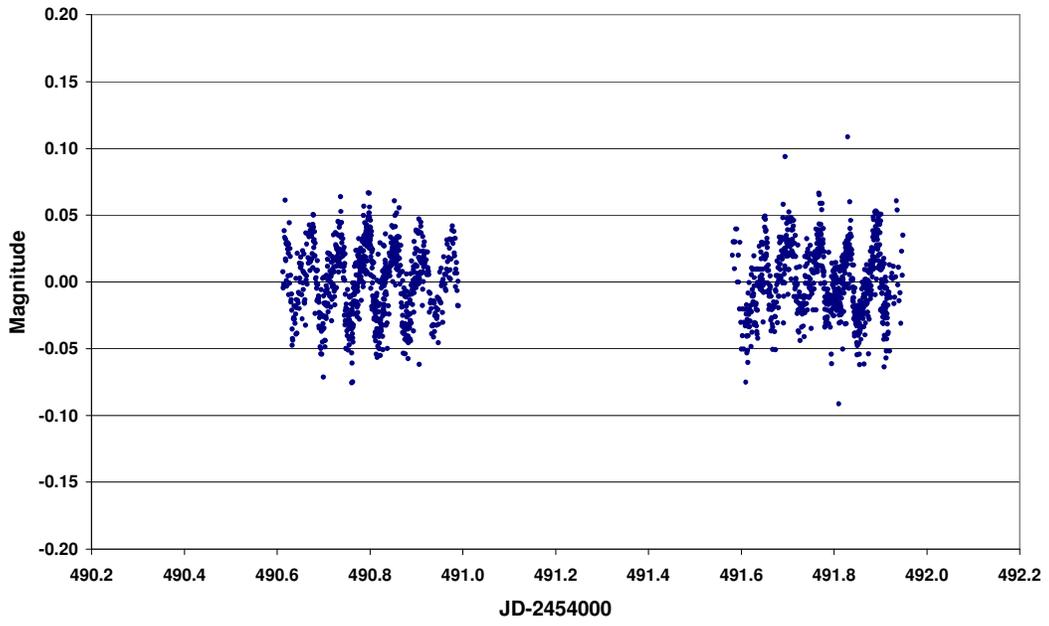

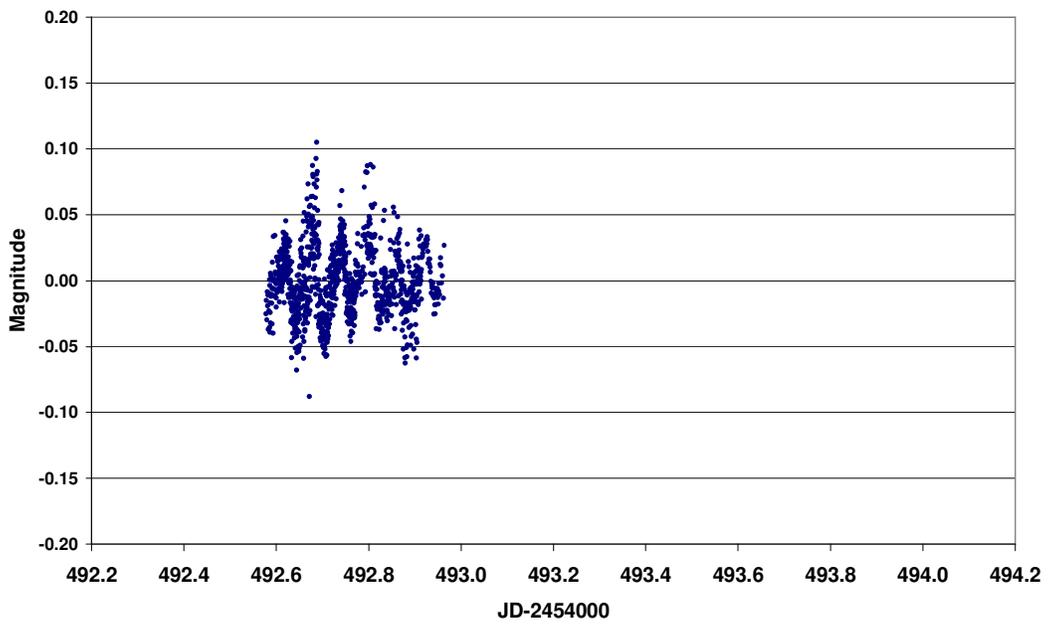

**Figure 3: De-trended light curve during the plateau phase**
*(note Figure 3 contains four light curves, the 2 above and the 2 on the previous page)*

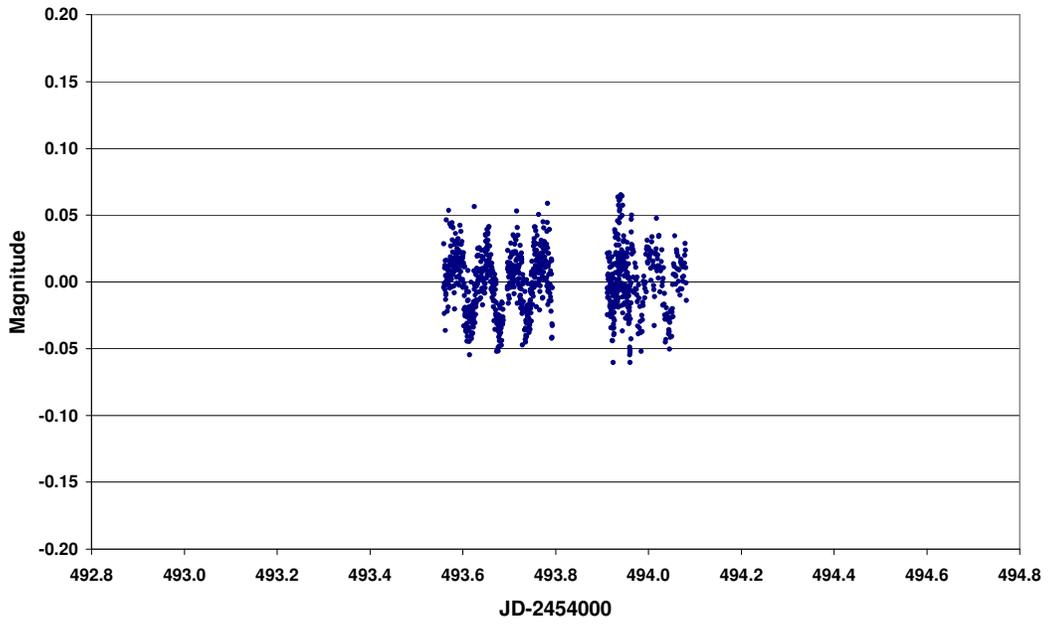

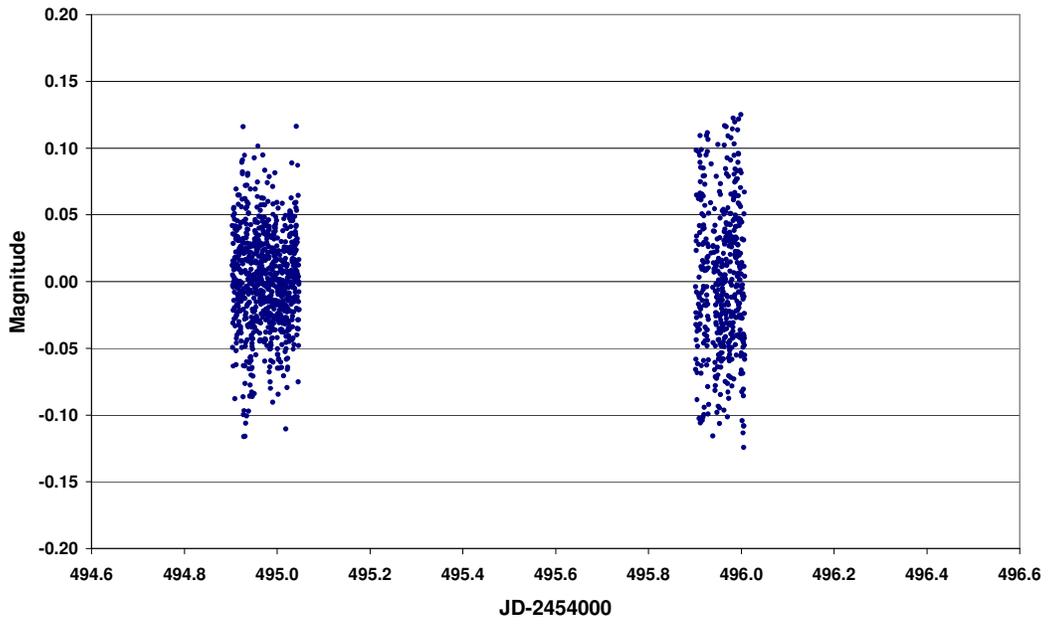

**Figure 4: De-trended light curve during the rapid decline**

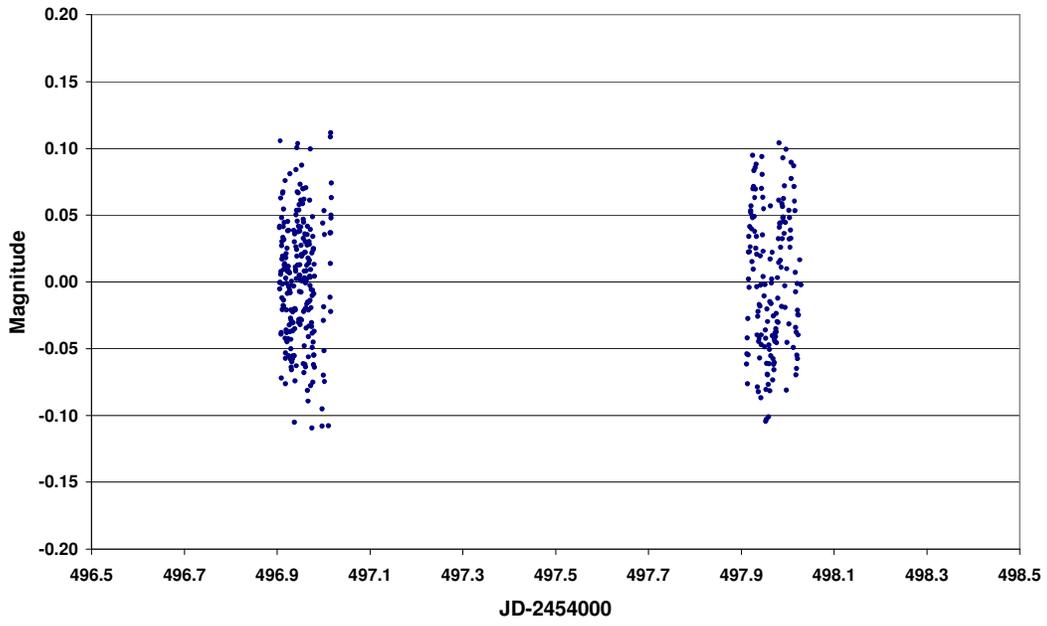

**Figure 5: De-trended light curve during the temporary minimum**

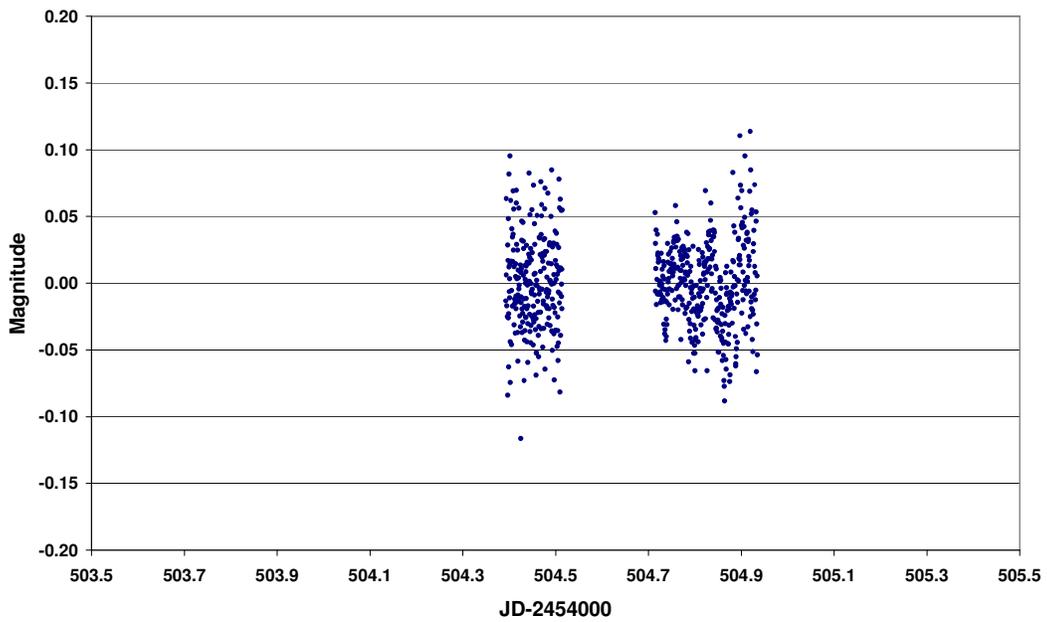

**Figure 6: De-trended light curve during the decline from the second echo outburst**

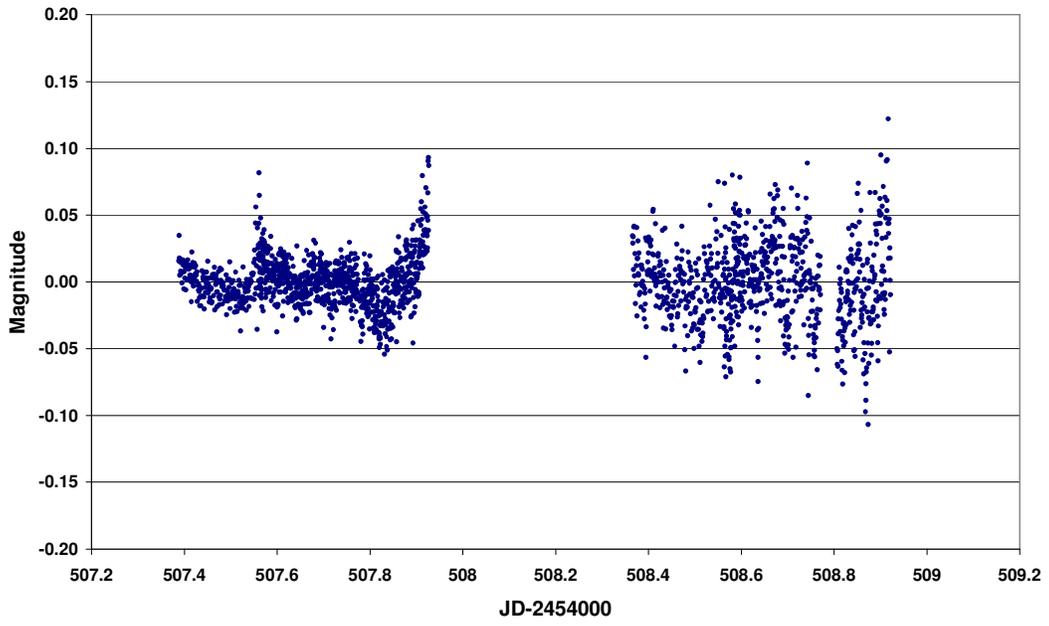

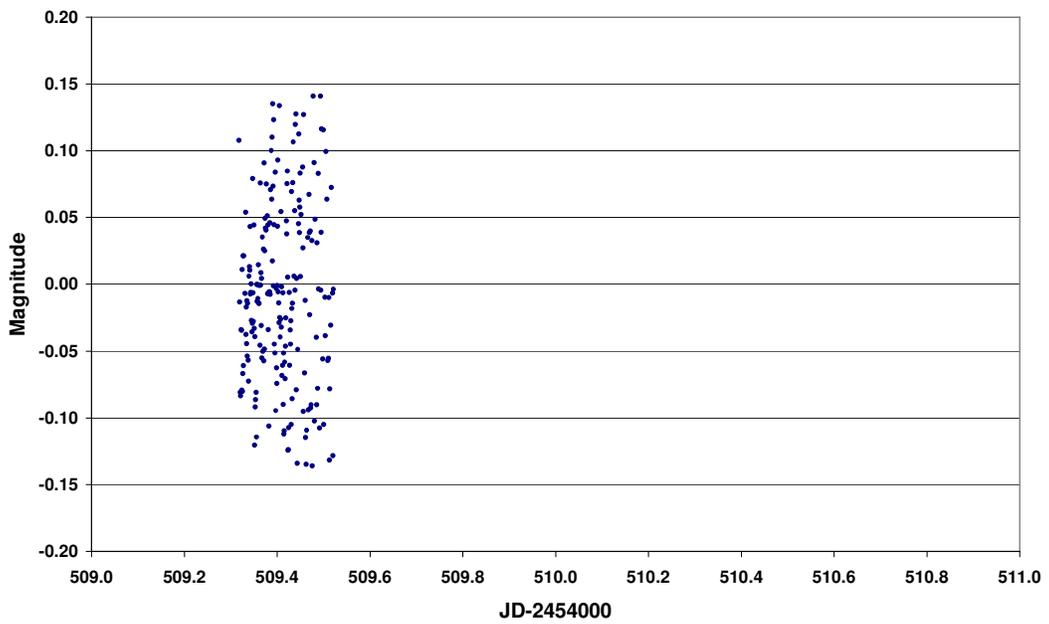

**Figure 7: De-trended light curve during the decline from the third echo outburst**

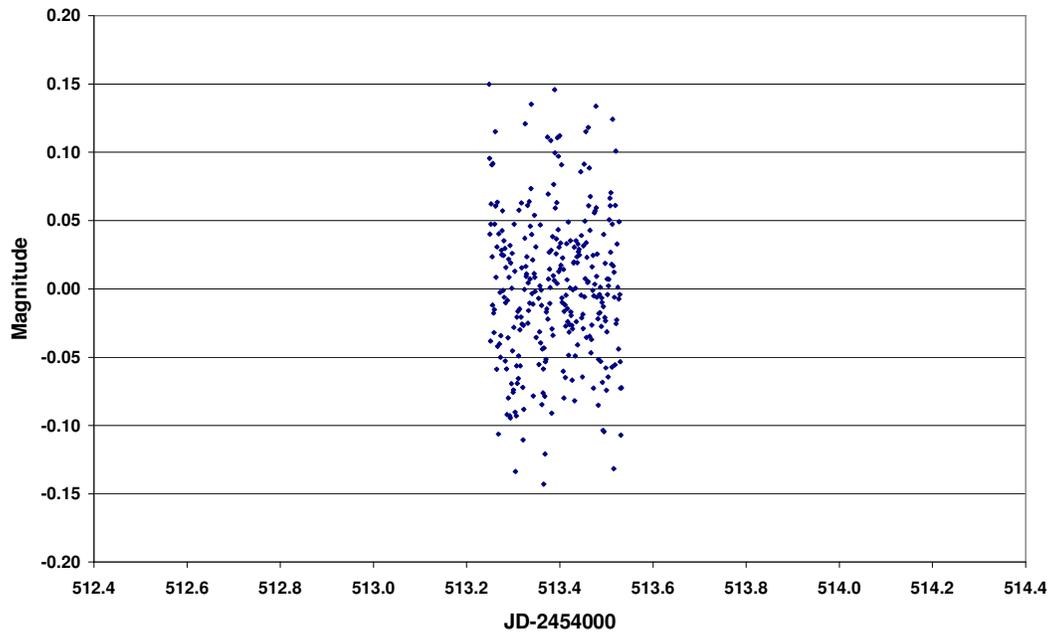

**Figure 8: De-trended light curve during the decline from the fourth echo outburst**

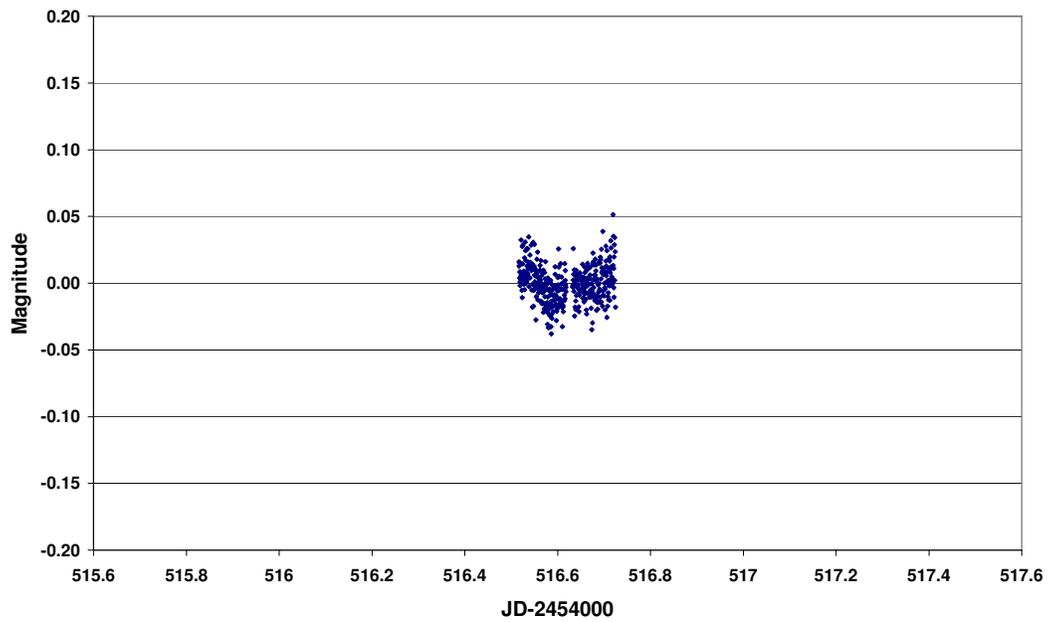

**Figure 9: De-trended light curve during the decline from the fifth echo outburst**

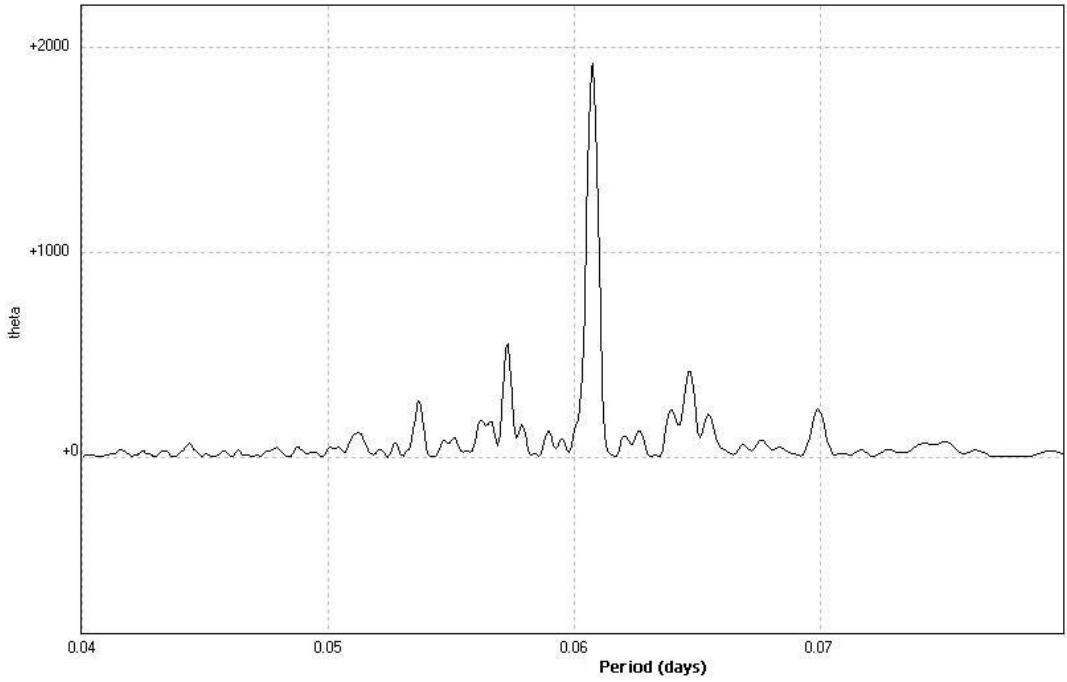

**Figure 10: Power spectrum of combined time-series data from the plateau phase**

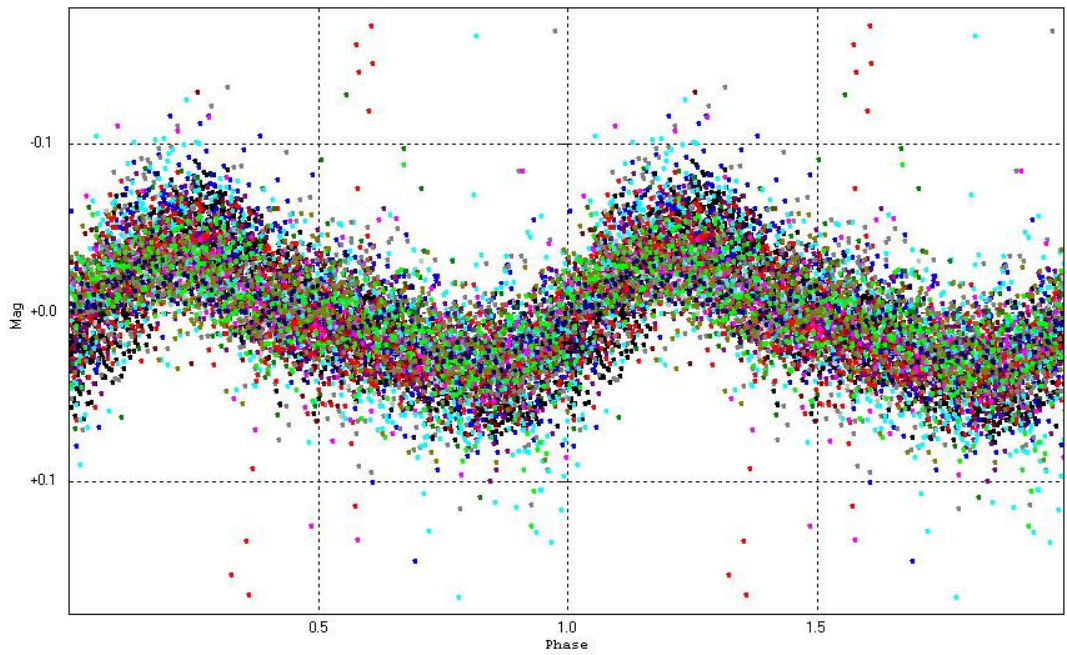

**Figure 11: Phase diagram from the plateau phase**